\documentclass[epj]{svjour}
\usepackage{graphicx}

\usepackage{color}
\usepackage{amssymb}
\usepackage{bbm}
\usepackage[intlimits]{amsmath}

\newcommand{\be}{\begin{equation}}
\newcommand{\ee}{\end{equation}}

\def\slashchar#1{\setbox0=\hbox{$#1$}  
   \dimen0=\wd0     
   \setbox1=\hbox{/} \dimen1=\wd1  
   \ifdim\dimen0>\dimen1   
      \rlap{\hbox to \dimen0{\hfil/\hfil}} 
      #1     
   \else     
      \rlap{\hbox to \dimen1{\hfil$#1$\hfil}} 
      /      
   \fi}

\makeatletter
\def\overbracket#1{\mathop{\vbox{\ialign{##\crcr\noalign{\kern3\p@}
\downbracketfill\crcr\noalign{\kern3\p@\nointerlineskip}
$\hfil\displaystyle{#1}\hfil$\crcr}}}\limits}
\def\underbracket#1{\mathop{\vtop{\ialign{##\crcr
$\hfil\displaystyle{#1}\hfil$\crcr\noalign{\kern3\p@\nointerlineskip}
\upbracketfill\crcr\noalign{\kern3\p@}}}}\limits}
\def\upbracketfill{$\m@th\makesm@sh{\llap{\vrule\@height3\p@\@width.7\p@}}%
\leaders\vrule\@height.7\p@\hfill
\makesm@sh{\rlap{\vrule\@height3\p@\@width.7\p@}}$}
\def\downbracketfill{$\m@th
\makesm@sh{\llap{\vrule\@height.7\p@\@depth2.3\p@\@width.7\p@}}%
\leaders\vrule\@height.7\p@\hfill
\makesm@sh{\rlap{\vrule\@height.7\p@\@depth2.3\p@\@width.7\p@}}$}
\makeatother

\begin{document}
\title{Heavy hybrid stars from multi-quark interactions}
\author{Sanjin Beni\' c\inst{1}
}                     
\institute{Department of Physics, Faculty of Science, University of Zagreb,\\
P.O.B. 331, HR-10002 Zagreb, Croatia} 
\date{Received: date / Revised version: date}
\abstract{
We explore the possibility of obtaining heavy hybrid stars within the
framework of the two flavor Nambu--Jona-Lasinio model that includes
8-quark interactions in the scalar and in the vector channel.
The main impact of the 8-quark scalar channel is to reduce the
onset of quark matter, while the 8-quark vector channel acts to stiffen
the equation of state at high densities.
Within the parameter space where the 4-quark vector channel is small,
and the 8-quark vector channel sizeable, stable stars with masses of
$2 M_\odot$ and above are found to hold quark matter in their cores.
} 
\maketitle

\section{Introduction}

One of the most severe constraints of 
the equation of state (EoS) of 
QCD at extreme
densities comes from the 
recent $2M_\odot$ mass determinations 
of PSR J1614-2230 \cite{Demorest:2010bx} and 
PSR J0348-0432 \cite{Antoniadis:2013pzd} having 
important consequences for the
existence of quark matter in compact 
stars \cite{Ozel:2006bv,Klahn:2006iw,Alford:2006vz,Lattimer:2010uk}.

In this work we assume compact stars are the so-called hybrid stars, composed
of a nuclear mantle and a quark core.
In order to obtain heavy hybrids stars, 
quark matter EoS should be stiff, with a low onset \cite{Alford:2013aca}.
This can be achieved in the MIT bag model 
by perturbative corrections to the EoS and small bag 
pressures \cite{Weissenborn:2011qu},
while within the Nambu--Jona-Lasinio (NJL) model 
large vector channel is used for stiffness \cite{Hanauske:2001nc},
while shift in the vacuum energy \cite{Pagliara:2007ph,Lenzi:2012xz} or 
introduction of superconductivity ensures low onset \cite{Klahn:2013kga}.
An influence on the maximum mass is 
expected by the nature of 
the phase transition itself \cite{Yasutake:2012dw},\cite{Masuda:2012ed,Alvarez-Castillo:2013spa}.

We propose an alternative scenario by taking into account 
the fact that NJL is a non-renormalizable 
effective model, valid 
up to some scale $\Lambda$.
As we approach this scale, say by increasing the 
chemical potential, higher dimensional 
operators should 
become important.
It has been pointed out \cite{Osipov:2006ev,Osipov:2007mk} that 
including higher scalar interactions can reduce the 
critical temperature
in the NJL model, bringing it in closer agreement with
lattice results at finite temperature \cite{Sakai:2009dv}.
For further work on higher dimensional operators in the context of
the NJL model
see \cite{Lee:1999fy,Mishustin:2003wq,Kashiwa:2006rc,Huguet:2006cm,Kashiwa:2007pc,Gatto:2010pt,Ohnishi:2011jv}.

Our aim is to make an initial study of 
the effect of multiquark interactions on occurrence 
of quark matter in heavy stars.
In this work we will use the NJL model parametrization of 
Ref.~\cite{Sakai:2009dv} where the critical temperature at
zero chemical potential is fitted
to lattice QCD.
In addition, we will also introduce the 8-quark vector channel,
as a natural candidate for playing a relevant
role at large densities reached in the cores
of compact stars.

Our main results are that with scalar 8-quark interactions
provided by Ref.~\cite{Sakai:2009dv} we are able 
to obtain stable hybrid
stars with small vector coupling in the 4-quark
and zero vector coupling in the 8-quark channel.
Second, we demonstrate that the mass of the star 
can be increased up to and above $2M_\odot$ with the 8-quark vector interaction, while
still keeping the 4-quark vector interaction low.

\section{NJL model with 8-quark interactions}
We work within the framework of a $N_f=2$ NJL model defined as follows
\be
\mathcal{L}=\bar{q}(i\slashchar{\partial}-m)q+\mu_q\bar{q}\gamma^0 q+
\mathcal{L}_4+\mathcal{L}_8~,
\label{eq:njllag}
\ee
where $\mu_q$ is the quark chemical potential and $m$ is 
the current mass.
The interaction terms are
\be
\mathcal{L}_4 =\frac{g_{20}}{\Lambda^2}
[(\bar{q}q)^2 + (\bar{q}i\gamma_5\boldsymbol{\tau} q)^2]-\frac{g_{02}}{\Lambda^2}(\bar{q}\gamma_\mu q)^2~,
\ee
\be
\begin{split}
\mathcal{L}_8 &= \frac{g_{40}}{\Lambda^8}[(\bar{q}q)^2 + 
(\bar{q}i\gamma_5\boldsymbol{\tau} q)^2]^2-
\frac{g_{04}}{\Lambda^8}(\bar{q}\gamma_\mu q)^4 \\ 
&-\frac{g_{22}}{\Lambda^8}(\bar{q}\gamma_\mu q)^2[(\bar{q}q)^2 + (\bar{q}i\gamma_5\boldsymbol{\tau} q)^2]~,
\end{split}
\label{eq:lag8}
\ee
where $\Lambda$ is the model cutoff.
The Lagrangian (\ref{eq:njllag}) may be
motivated by the large $N_c$ counting: while the 4-quark
couplings scale as $1/N_c$, the 8-quark couplings 
scale as $1/N_c^3$ \cite{Alkofer:1990uh}.

In the mean-field approximation the Lagrangian becomes
\be
\mathcal{L}_\mathrm{MF} = \bar{q}(i\slashchar{\partial}-M)q+\tilde{\mu}_q\bar{q}\gamma^0 q-U~, 
\ee
where
\be
M=m+2\frac{g_{20}}{\Lambda^2}\langle\bar{q}q\rangle+4\frac{g_{40}}{\Lambda^8}\langle\bar{q}q\rangle^3
-2\frac{g_{22}}{\Lambda^8}\langle\bar{q}q\rangle\langle q^\dag q\rangle^2~,
\ee
\be
\tilde{\mu}_q = \mu_q - 2\frac{g_{02}}{\Lambda^2}\langle q^\dag q\rangle-
4\frac{g_{04}}{\Lambda^8}\langle q^\dag q\rangle^3-2\frac{g_{22}}{\Lambda^8}\langle\bar{q}q\rangle^2
\langle q^\dag q\rangle~,
\label{eq:consmass}
\ee
and the classical potential
\be
\begin{split}
U &= \frac{g_{20}}{\Lambda^2}\langle\bar{q}q\rangle^2 + 
3 \frac{g_{40}}{\Lambda^8}\langle\bar{q}q\rangle^4-
3\frac{g_{22}}{\Lambda^8}\langle\bar{q}q\rangle^2
\langle q^\dag q\rangle^2\\
&-\frac{g_{02}}{\Lambda^2}\langle q^\dag q\rangle^2
-3\frac{g_{04}}{\Lambda^8}\langle q^\dag q\rangle^4 ~.
\end{split}
\ee

Integrating out the quark degrees of freedom, the 
full thermodynamic potential takes the following form
\be
\begin{split}
\Omega &= U -2 N_f N_c\int \frac{d^3 p}{(2\pi)^3}
\Bigl\{E + T\log[1+e^{-\beta(E-\tilde{\mu}_q)}]\\
&+ T\log[1+e^{-\beta(E+\tilde{\mu}_q)}]\Bigr\} + \Omega_0~,
\end{split}
\label{eq:pot}
\ee
where $E = \sqrt{\mathbf{p}^2+M^2}$ and $\beta=1/T$ and where
$\Omega_0$ ensures zero pressure in the vacuum.
The model is solved by minimizing the thermodynamic potential
with respect to the mean-fields 
$X=\langle\bar{q}q\rangle, \langle q^\dag q\rangle$, i. e.
\be
\frac{\partial \Omega}{\partial X} = 0~. 
\ee

In this work we use the parameter set of 
Ref.~\cite{Sakai:2009dv}
$g_{20} = 1.864$, $g_{40} = 11.435$, $m=5.5$ MeV $\Lambda=631.5$ MeV.
The vector channel strengths are treated as free parameters,
quantified by
\be
\eta_2 = \frac{g_{02}}{g_{20}} \, , \qquad 
\eta_4 = \frac{g_{04}}{g_{40}}~.
\ee
We are interested in a particular region of the
vector channel couplings where the $g_{02}$ coupling is kept
small, while $g_{04}$ is increased.
The reason behind our choice is as following.
Since heavy hybrid stars require a stiff EoS, a repulsive 
vector coupling should be present.
As the vector coupling renormalizes
the chemical potential, it delays the onset of quark matter.
This leads to a scenario where the hadronic mantle becomes too large
for the pressure in the quark core to be able to hold the star 
against gravitational collapse.
Therefore, the appearance of quark matter in such a
scenario usually makes the star unstable.
An attractive channel like e. g. superconductivity needs to be invoked 
in order to lower the onset \cite{Klahn:2013kga}.
We point out that an alternative 
microscopic picture is possible
with multiquark interactions:
whereas stiffening of the EoS is provided 
by the 8-quark vector channel interaction,
lowering of the onset is accomplished 
by introducing a sizeable 8-quark
scalar channel interaction and keeping the 4-quark 
vector channel interaction
small.

Our choice of the phenomenologically 
interesting parameter space will have an effect of stiffening
the quark matter at higher densities, while at the same
time keeping the transition density low.
Due to this restriction 
we will also put $g_{22}=0$ by hand.
Namely, the operator controlled by the size of $g_{22}$ can
be important at moderate $\mu_q$ only if the vector mean-field
is sizeable, which will not be the case for the 
parameter region of small $g_{02}$ in which we are interested in.
In addition, it will not be important at high $\mu_q$ since there
the scalar mean-field is zero.

\section{Equation of state}

\begin{figure}[t]
\centering
\begin{tabular}{cc}
\includegraphics[scale=0.35,clip]{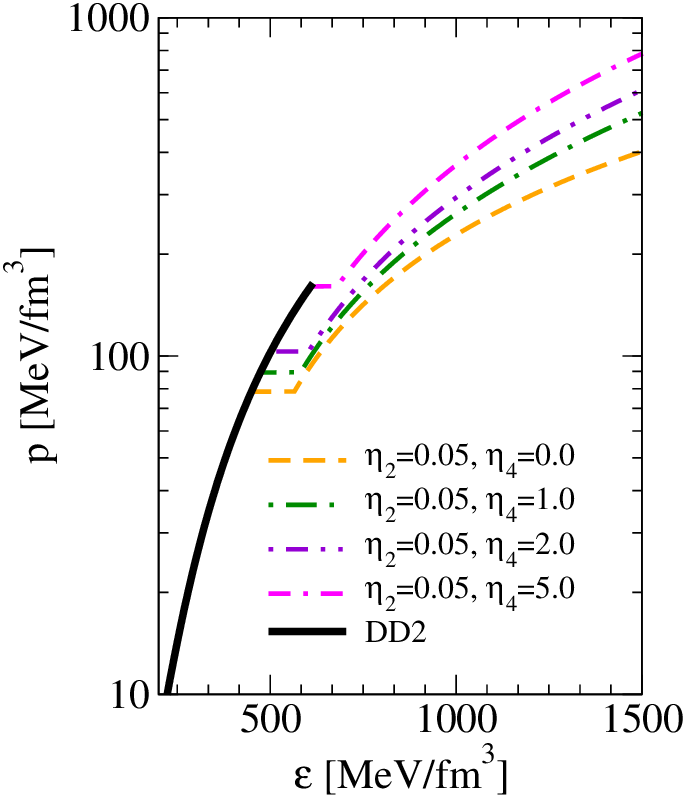}&
\includegraphics[scale=0.35,clip]{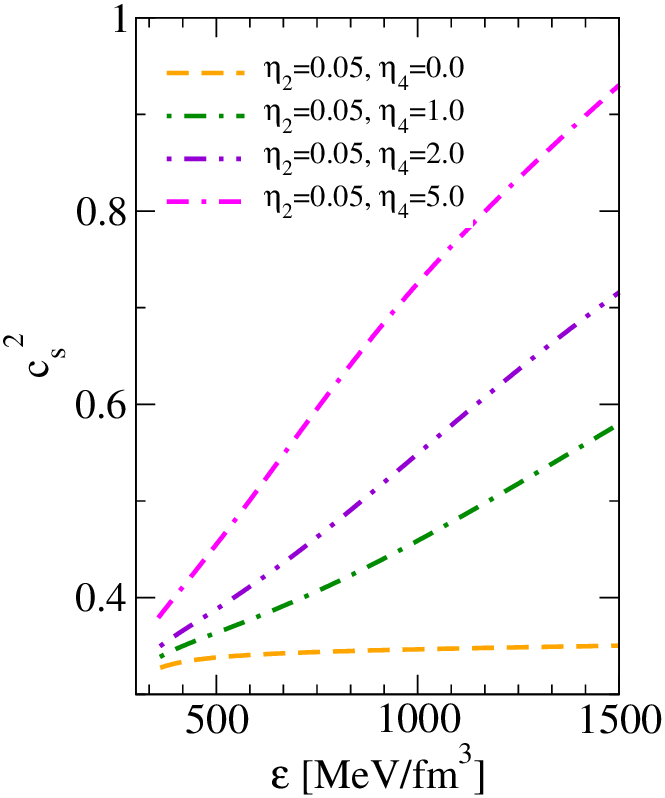}\\
\end{tabular}
\caption{(Color online) On the left panel we show
the EoS of hybrid matter.
Thick black curve is the hadronic contribution.
Flat region corresponds to Maxwell construction.
We fix $\eta_2=0.05$.
Quark matter curves are denoted as following:
the orange, dashed line accounts for $\eta_4=0.0$, dash-dotted,
green line for $\eta_4=1.0$, dash-double-dotted, indigo line for $\eta_4=2.0$ and double-dash-dotted, magenta 
line for $\eta_4=5.0$.
Right panel shows the speed of sound of quark 
matter with the same labeling.
}
\label{fig:eos1}
\end{figure}
By evaluating the full thermodynamical potential (\ref{eq:pot}) at 
the minimum, we obtain the quark matter EoS
as $p_q = -\Omega$.
The quark number density $n_q$ and energy 
density $\epsilon_q$ are defined 
as follows
\be
n_q=-\frac{\partial p_q}{\partial \mu_q}\, , \qquad
\epsilon_q = -p_q+n_q\mu_q~.
\ee
Beta equilibrium is taken into account by the weak processes
$d\to u+e^- +\bar{\nu}_e$, $\mu^- \to e^- + \nu_\mu +\bar{\nu}_e$,
implying 
$\mu_u = \mu_d + \mu_e$, $\mu_\mu=\mu_e$,
where the neutrino chemical potential is set to zero.
Finally, the baryon chemical potential, and the baryon density
are $\mu_B = 3\mu_q = 2\mu_d+\mu_u$ and $n_B=n_q/3$, respectively.

For the nuclear matter we choose the 
DD2 EoS \cite{Typel:1999yq,Typel:2009sy}.
The transition from nuclear to quark matter is provided by
the traditional Maxwell construction.
Therefore, the EoS is obtained by requiring local 
charge neutrality.
The full pressure in the quark phase takes into account
the contribution of electrons and muons
\be
p(\mu_q) = p_u(\mu_u)+p_{d}(\mu_d) + p_e(\mu_e)+p_\mu(\mu_\mu)~,
\ee
where $p_{e,\mu}$ is the pressure of a free electron (muon) gas.
We emphasize that the Maxwell 
construction is merely the limit of large surface tension 
at the quark-hadron interface in a more elaborate approach that takes into
account finite size effects, see \cite{Maruyama:2007ey}.
The following results ultimately depend also 
on the choice of the construction of
the phase transition.

The hybrid EoS are given on Fig.~\ref{fig:eos1} for $\eta_2=0.05$ 
and a range of $\eta_4$.
Owing to 8-quark scalar interactions, and small 4-quark vector 
interactions, onset of quark matter 
is rather low: for $\eta_2=0.05$ and $\eta_4=0.0$
it is around $\epsilon\simeq 500$ MeV/fm$^3$.
Even a drastic increase of $\eta_4$ has 
barely any influence on the onset: this is only natural, since there
is extra suppression due to high dimensionality of the corresponding
operator.

The influence of $\eta_4$ is best seen by inspecting
the speed of sound: while small or almost vanishing 
vector interactions
yield the relativistic value 
$c_s^2 \simeq 1/3$,
the speed of sound significantly increases with $\epsilon$
already for $\eta_4=1.0$, see right 
panels of Fig.
~\ref{fig:eos1}.
This gradual stiffening of the EoS can also
be seen as a microscopic mechanism
of a postulated scenario of a medium-dependent 
parameter $\eta_2$
\cite{Blaschke:2013ana}, which is able to provide
a very stiff quark EoS \cite{Alford:2013aca}. 
Assuming that the vector mean-field
is given by the density of massless fermions, an approximative 
expansion in $g_{02}$ and $g_{04}$ for the speed of sound 
$c_s^2=\partial p/\partial \epsilon$ as a function of
the quark chemical potential can be shown to hold
\be
\begin{split}
c_s^2 &\simeq \frac{1}{3}+\frac{32g_{02}}{6\pi^2}\frac{\mu_q^2}{\Lambda^2}+
\frac{512g_{02}^2}{4\pi^4}\frac{\mu_q^4}{\Lambda^4}+
\frac{75776 g_{02}^3}{24\pi^6}\frac{\mu_q^6}{\Lambda^6}\\
&+\frac{3768320 g_{02}^4}{48\pi^8}\frac{\mu_q^8}{\Lambda^8}+
\frac{8192g_{04}}{48\pi^8}\frac{\mu_q^8}{\Lambda^8}~,
\end{split}
\label{eq:sos}
\ee
illustrating that the $g_{04}$ term starts to be 
important only at higher chemical potentials.
While the 4-quark vector interaction respects causality, strong
8-quark vector interaction may violate the causal limit, see
Ref.~\cite{Burvenich:2003mv} for an 
explicit example in the nucleonic NJL model.
This is the reason why the EoS with $\eta_4=5.0$ turns
acausal already at $\epsilon\sim 2000$ MeV/fm$^3$.
In addition, for such high densities, the quark chemical 
potential is approaching $\Lambda$ where the model should not be
trusted anymore.
Therefore, results obtained for this extreme 
scenario are given for illustrative purposes.

\section{Hybrid stars}

\begin{figure}[htb]
\centering
\includegraphics[scale=0.3]{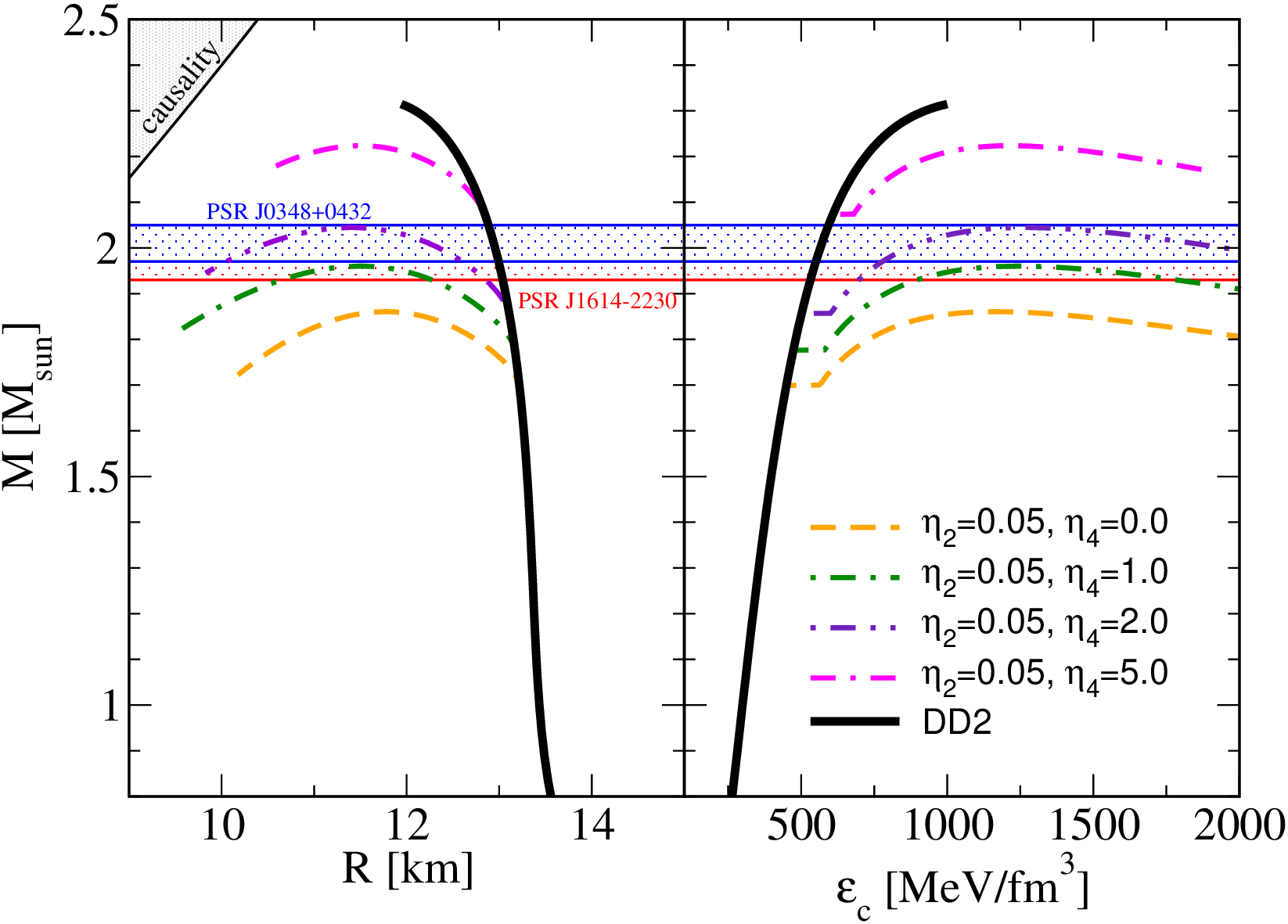}
\caption{(Color online) Left panel shows the $M-R$
and the right panel the $M-\epsilon_c$ diagram of compact
star solutions within the model. 
We fix $\eta_2=0.05$.
The detachment from the hadronic branch given by the 
thick, black curve and the maximum mass depends on $\eta_4$.
Hybrid star branches appear in the same line styles as the corresponding
EoS on Fig.~\ref{fig:eos1}.
Hatched regions mark experimental constraints from two heaviest 
sources,
lower, red is \cite{Demorest:2010bx}, while higher, blue 
is \cite{Antoniadis:2013pzd}.}
\label{fig:mrec1}
\end{figure}
The static, spherically symmetric stars
are obtained as solutions of the Tolman-Oppenheimer-Volkoff (TOV)
equations for the EoS shown in the previous section.
On Fig.~\ref{fig:mrec1} we display 
the resulting mass-radius and mass-central
energy density diagrams 
for $\eta_2=0.05$.

Due to the early onset of quark matter, on 
Fig.~\ref{fig:mrec1} we are able to obtain
stable stars with pure quark matter in their cores 
even for very small vector coupling.
Such a scenario is not easy to achieve in the NJL
model with only fourth order scalar and 
vector operators \cite{Klahn:2013kga,Orsaria:2013hna}, see 
however \cite{Lenzi:2012xz}.
The strong vector interaction increases the speed of sound
and makes quark matter stiff, but at the same time it 
appears at too high energy densities.
Higher dimensional vector operator stiffens the EoS
and without influencing the onset significantly,
gives a mechanism for $2 M_\odot$ hybrid stars, see Fig.~\ref{fig:mrec1}.
In order to cover the present experimental window provided
by PSR J1614-2230 and PSR J0348-0432 we have found that
values $\eta_4$ up to $\eta_4\sim 2.0$ are sufficient.
The extreme case with $\eta_4=5.0$ leads to a 
significant increase in the mass, yielding $M\sim 2.25 M_\odot$.

\section{Conclusions}

Observing heavy compact stars offers a promising 
perspective on constraining
the cold, dense EoS beyond saturation density.
A compelling discrimination of many possible scenarios of dense matter
will be possible once precise measurements of also star
radii become available \cite{Miller:2013tca}.
The data of both masses and radii will enable Bayesian inversion
of the TOV equation leading to the EoS \cite{Steiner:2012xt}.

We have studied one possible scenario
where multi-quark interactions coming from higher 
dimensional operators
in the NJL model might play an 
important role at large densities.
A sizeable 8-quark scalar channel is introduced in order to
achieve a low onset.
With a small 4-quark vector coupling onset is still low, while 
the relatively large 8-quark coupling
is used to gradually stiffen the EoS at high densities.
Within this parameter space we were able to fulfill
and go beyond the $2M_\odot$ constraint.

The scenario of small 4-quark vector coupling is in accordance with the
results of Ref.~\cite{Steinheimer:2010sp,Schramm:2013rya}, but a more thorough 
study is needed to reveal how
would the introduction of the 8-quark vector coupling influence their results.
In addition, we stress that the consensus concerning the status 
of vector interactions in quark
matter is yet to be reached in the community, as some 
other studies \cite{Sakai:2009dv,Bratovic:2012qs,Contrera:2012wj,Sugano:2014pxa} advocate 
a different scenario where the 4-quark vector coupling is sizeable. 
Using the results from the lattice QCD measurements at finite $T$ 
\cite{Cheng:2008zh,Kaczmarek:2011zz,Bazavov:2011nk} and 
at imaginary chemical potential \cite{de Forcrand:2002ci} a 
more thorough study of the model
presented here is needed to achieve better 
constraints on the three vector
channel couplings, and to investigate the consequences 
at large densities.  
Finally, it remains to be explored what can such a setup
say about the existence of strangeness in compact stars.

\subsection*{Acknowledgments}
We acknowledge illuminating discussions 
with D.~Blaschke 
and thank D.~E.~Alvarez-Castillo for help with the TOV code.
The author thanks the Yukawa Institute for Theoretical Physics, Kyoto 
University, where part of this work was performed during the YITP-T-13-05 
workshop on ``New Frontiers in QCD''.
This work is supported by the
MIAU project of the Croatian Science Foundation and 
by the the COST Action
MP1304 NewCompStar.


\end{document}